%% file: main.tex
\documentclass[runningheads]{llncs}

\include{macros}

\newacronym{CRT}{CRT}{Chinese Remainder Theorem}
\newacronym{HE}{HE}{Homomorphic Encryption}
\newacronym{LSB}{LSB}{Least Significant Bit}
\newacronym{MSB}{MSB}{Most Significant Bit}
\newacronym{SIMD}{SIMD}{Single Instruction Multiple Data}
\newacronym{RLWE}{R-LWE}{Ring Learning With Errors}
\newacronym{ML}{ML}{Machine Learning}
\newacronym{PPML}{PPML}{Privacy-Preserving Machine Learning}
\newacronym{ResNet}{ResNet}{Residual Net}
\newacronym{SOTA}{SOTA}{State-Of-The-Art}
\newacronym{SC}{SC}{skip connection}
\newacronym{DL}{DL}{Deep Learning}
\newacronym{CNN}{CNN}{Convolutional Neural Network}
\newacronym{DNN}{DNN}{Deep Neural Network}
\newacronym{MLaaS}{MLaaS}{ML as a Service}

\begin{document}

%\input{_header}
%\thispagestyle{empty} 
%\input{_abstract}

%~
%\newline
%\noindent\textbf{Cover page: short academic paper.}

\input{_header}

\pagenumbering{arabic} 
\input{_abstract}

\section{Introduction}

The use of \gls{HE} to construct \gls{PPML} solutions, e.g., secure \gls{DNN} inference on the cloud, becomes more and more realistic. For example, Gartner \cite{gartner} predicted that in 2025, $50\%$ of large enterprises will adopt HE-based solutions. In addition, we see many companies and academic institutes collaborate in global activities such as HEBench \cite{hebench} and the \gls{HE} standardization efforts \cite{standard}. The main reason is, of course, that \gls{HE} allows finance and health organizations to comply with regulations such as GDPR \cite{GDPR} and HIPAA \cite{HIPAA} when uploading sensitive data to the cloud. 

One principle scenario of \gls{HE}-based \gls{PPML} solutions involves two entities: a user and a semi-honest cloud server that performs \gls{ML} computation on HE-encrypted data. Specifically, the cloud offers an \gls{MLaaS} solution, where it first trains a model in the clear, e.g., a \gls{DNN}, and then, uses it for inference operations on the clients' data. On the other side, the client first generates its own \gls{HE} keys, stores the secret key, and uploads the public and evaluation keys to the cloud. Subsequently, upon demand, it encrypts secret samples and submits them to the cloud that uses the client's public and evaluation keys to perform the model inference operation. The final encrypted results are sent back to the client who decrypts them using its private key. 

The clients' data is kept confidential from the server during the entire protocol due to HE, while the cloud model is never sent to the client, which allows the cloud to monetize its \gls{MLaaS} service. In this paper, we focus on this scenario but stress that our study can be used almost without changes in many other threat models.

One downside of \gls{HE}-based solutions is their latency. While there are many software and hardware improvement that make HE-based solutions practical such as \cite{helayers, pmlr-v162-lee22e}, there is still a gap between computing on encrypted data and computing on cleartexts, where our goal is to reduce that gap. Our starting point is a recent study \cite{e2e} that pointed out on skip connections as a major contributor to the overall latency of secure inference solutions that use \glspl{DNN}. The authors of \cite{e2e} suggested removing the skip connection at the cost of some accuracy degradation or replacing the skip connections using several heuristics. We continue this line of work by suggesting using modern techniques that allow training \glspl{DNN} while maintaining good accuracy. Specifically, 
we replace mid-term skip connections in \glspl{DNN} with short-term (Dirac parameterization) \cite{diracnet} and long-term (Shared source skip connection) \cite{new-highway}.
 
\paragraph{Our contribution} We used ResNet50, a state-of-the-art network in terms of size that can run efficiently under HE \cite{e2e, pmlr-v162-lee22e}, as our baseline. We modified it to be HE-friendly, a term that we explain later, and apply the above techniques. Our experiments show that using this approach we were able to reduce the number of HE bootstrap operations by $\times 1.36 - 1.75$ and thus the overall CPU time by x1.3.

\paragraph{Organization.}
The document is organized as follows. In section \ref{sec:he} we provide some background about HE. We describe skip connections and their variants in Section \ref{sec:sc}. Our experiments and results are presented in Section \ref{sec:exp} and we conclude in Section \ref{sec:conc}.

\section{\acrfull{HE}}\label{sec:he}
\gls{HE} is an encryption scheme that encrypts input plaintext from a ring $\R_1(+, *)$ into ciphertexts in another ring $\R_2(\oplus, \odot)$, i.e., it contains the encryption function $\Enc:\R_1 \rightarrow \R_2$ and decryption function $\Dec:\R_2 \rightarrow \R_1$, where a scheme is correct if $\Dec(\Enc(x)) = x$. In addition, HE schemes include homomorphic addition and multiplication operations such that
$\Dec(\Enc(x) \oplus \Enc(y)) = x + y$ and $\Dec(\Enc(x) \odot \Enc(y))  = x * y$ see survey in \cite{Halevi2017}. In our experiment, we use CKKS \cite{ckks2017, ckks-rns} an approximately correct scheme, i.e., for some small $\epsilon > 0$ that is determined by the key, it follows that $|x - \Dec(\Enc(x))| \le \epsilon$ and the same modification applies to the other equations.

\paragraph{Chain Index and Bootstrapping.}
HE ciphertexts and particularly CKKS ciphertexts have a limit on the number of multiplications they can be involved with before a costly bootstrap operation is required. To this end, every ciphertext includes an extra metadata parameter called the "multiplication chain index'' (a.k.a. modulus chain index) or \ChainIdx. Ciphertexts start with a \ChainIdx of $0$ and after every multiplication of two ciphertexts with \ChainIdx of $x$ and $y$, the result has a \ChainIdx of $\max(x,y) + 1$, where at least a \ReScale operation is required. This process continues until the ciphertext reaches the predefined limit, which was originally set by the client to achieve the desired level of security and performance.  To enable further computation on a ciphertext, a $\Bootstrap$ operation is performed to reduce its \ChainIdx, or even reset it back to $0$. In general, many HE-based applications attempt to minimize the number of $\Bootstrap$ invocations and this is also our goal in this paper.

There are two options for adding or multiplying two ciphertexts $c_1$, $c_2$ with $\ChainIdx=x,y$, respectively, where w.l.o.g $x > y$: a) adjust $c_1$ to have a $\ChainIdx=y$ by invoking $\ReScale(\Bootstrap(c_1), y)$; or b) invoke $\ReScale(c_2, x)$. This first option is costlier because it invokes both \ReScale and \Bootstrap in advance while the other approach leaves the bootstrap handling to future operations. However, this approach is preferred when $c_1$ is expected to be added to multiple ciphertexts with lower chain indices. In that case, we perform only one \Bootstrap operation on $c_1$ instead of many on the other operations' results. An automatic bootstrapping placement mechanism is expected to consider the above. 

\paragraph{HE Packing.} Some \gls{HE} schemes, such as CKKS \cite{ckks-rns}, operate on ciphertexts in a homomorphic \gls{SIMD} fashion. This means that a single ciphertext encrypts a fixed-size vector, and the homomorphic operations on the ciphertext are performed slot-wise on the elements of the plaintext vector. To utilize the SIMD feature, we need to pack and encrypt more than one input element in every ciphertext. The packing method can significantly impact both bandwidth, latency, and memory requirements. In this paper we decided to rely on IBM HElayers, which provides efficient packing capabilities for \glspl{DNN} through the use of a new data structure called tile tensors \cite{helayers}. We stress that adding or multiplying two ciphertexts that represent different tile tensor shape is problematic and an extra transformation is needed. While automatically handled by HElayers, one of goals is to also save these transformations.

\section{Skip connections}\label{sec:sc}
Skip connections, a.k.a, residual connections \cite{sm}, are crucial components in modern network architectures. Given several layers f(x), applying skip-connection to the layer means wrapping f(x) with a function $S_f(x) = f(x) + x$. For real-world applications, networks without skip connections are hard to train, especially very deep networks. Skip connection solves optimization issues such as (i) vanishing gradients (ii) exploding gradients \cite{sp2}, or (iii) shattering gradients \cite{SP3}. In practice, modern architectures heavily rely on skip connections e.g., Transforms~\cite{vaswani2017attention} ViT~\cite{dosovitskiy2020image}, LLMs, CNNs~\cite{liu2022convnet}, WaveNet~\cite{oord2016wavenet}, GPT~\cite{GPT}, and \gls{ResNet} who has became one of the most cited \gls{DNN} of the 21st century.
When considering cleartext networks, skip-connections require only a simple addition, and thus provide an efficient solution that enables easier optimization of \glspl{DNN}. Moreover, they also play a fundamental role in modern \gls{DL} solutions.
While skip-less networks exist, and new variants appear from time to time e.g., \cite{NS1, NS2}, they are rarely used in real-world applications, as they tend to perform poorly in complex scenarios and noisy data.

\subsection{Handling skip connections in HE} 

Observation 3.1 of \cite{e2e} explains the relation between skip connections and bootstrapping operations.

\begin{observation}[observation 3.1 \cite{e2e}]\label{obs:sc}
Given a \acrlong{SC} layer $S_f(x) = x + f(x)$, where $f$ is a combination of some layers. When running under HE,
\begin{enumerate}
    \item $\ChainIdx\left(S_f(x)\right) \in \{\ChainIdx(x), \ChainIdx(f(x))\}$.
    
    \item When $\ChainIdx(x) \neq \ChainIdx(f(x))$ the skip connection implementation invokes either a \ReScale or a  \Bootstrap operation and may increase the overall multiplication depth of the network by $|\ChainIdx(x) - \ChainIdx(f(x))|$.
    \end{enumerate} 
\end{observation}

In addition, the authors of \cite{e2e} explained that the cost of $S_f(x)$ can be even higher because the input $x$ or $f(x)$ may need to go through some transformations before adding them together, which is the case with the HElayers SDK. Given the latency costs associated with implementing skip connections under HE, \cite{e2e} proposed either removing skip connections by first training a network and then gradually removing the connections, which resulted in some accuracy degradation, or that they suggested some heuristics to reroute these connections, which offers a tradeoff between latency and accuracy. Here, we suggest another heuristic that brings knowledge from the AI domain into the HE domain. Particularly, we replace \glspl{DNN} skip connections with Dirac parameterization \cite{diracnet} and shared-source skip connection \cite{new-highway}. Informally speaking, shared-source skip connections connect the output of the initial layer or input with the output of different locations in the network. The reason this reduces the number of bootstraps is that after the initial layer, the chain index is very low. In addition, we can aim to add these connections only to layer outputs that share the same tile tensor shape and thus save reshape operations. Dirac parameterization is explained as follows: let $\sigma(x)$ be a function that combines non-linearity and batch normalization, then a standard convolutional layer in ResNet is of an explicit form $y = x + \sigma(W \odot x)$, where a Dirac parameterization is of the form $y = \sigma (diag(a)I + W) \odot x = \sigma(diag(a)x + W \odot x)$. This addition, helps training and does not affect the latency of the secure inference.

\begin{figure}[t!]
    \centering
    \includegraphics[width=0.8\textwidth]{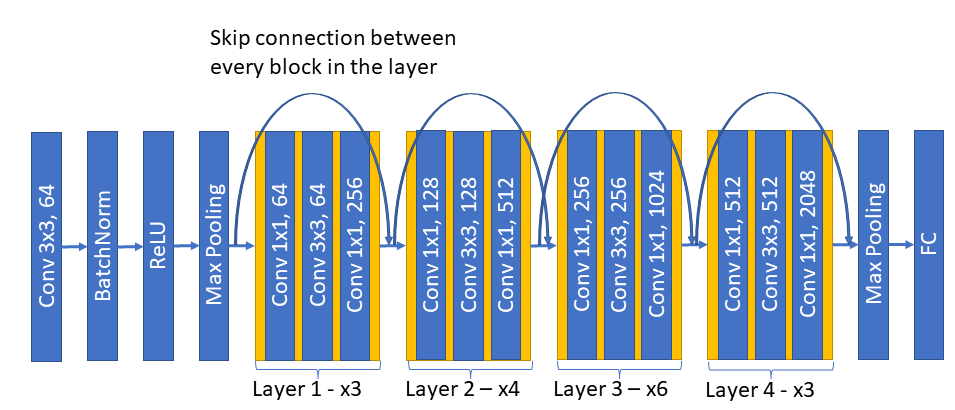}
    \caption{An illustration of ResNet50 every layer contains several blocks and there is a skip connection between every block.}
    \label{fig:resnet}
\end{figure}
\begin{figure}[t!]
    \centering
    \includegraphics[width=0.8\textwidth]{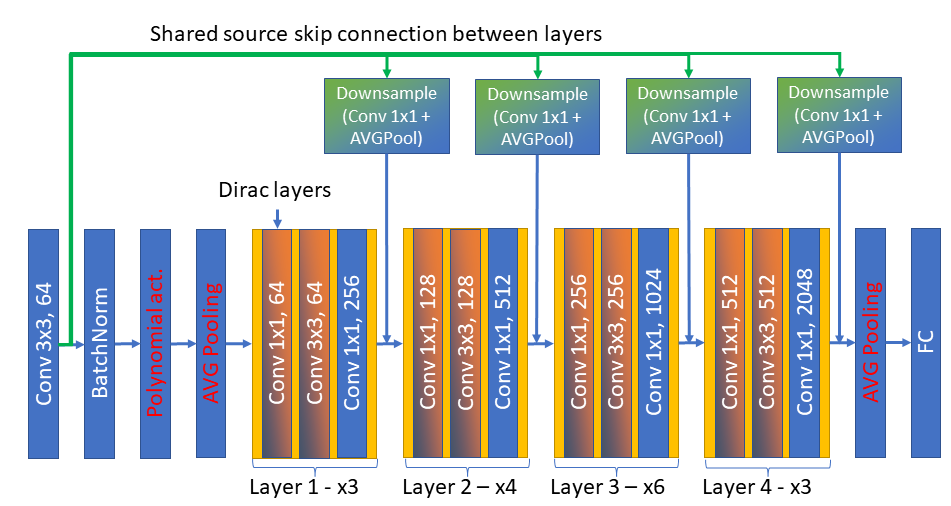}
    \caption{An illustration of our modified HE-friendly ResNet50. Every layer contains several blocks and there is a shared-source skip connection from the first layer output to the output of the four other layers. Red layers were modified to make the network HE-friendly as in \cite{CryptoNets2016}.}
    \label{fig:modresnet}
\end{figure}

Figure \ref{fig:resnet} shows a standard \gls{ResNet} network with skip connection in their original places. In contrast, Figure \ref{fig:modresnet} shows our modification. Because HE only supports polynomial operations, we first replace non-polynomial layers with polynomial layers (red font layers) to achieve a HE-friendly network. Specifically, we replace MaxPool with AVGPooling and ReLU activations with polynomial activations similar to \cite{CryptoNets2016}. Subsequently, we removed all mid-term skip connections and added long-term shared-source connections (green arrows) from the output of the first convolution layer to every one of the four layers' outputs. To ensure that the dimensions match we added $1 \times 1$ convolutional and average pooling layers to these connections. Note that these layers can be performed on the server side but also by the client if we consider a split network, where the first layer is performed on the client side. This offers a tradeoff between latency and bandwidth. Finally, we added low-term Dirac parameterization to the first two convolution layers of every block, where the stride is $1$ (orange blocks).

\section{Experiments}\label{sec:exp}

In our experiments, we used a single NVIDIA A100-SXM4-40GB GPU with 40GB of memory for training and for secure inference an Intel\circler Xeon\circler CPU E5-2699 v4 @ 2.20GHz machine with 44 cores (88 threads) and 750GB memory. In addition, for inference we used HElayers~\cite{helayers} version $1.5.2$, where we set the underlying HE library to HEaaN\cite{heaanCode} targeting $128$ bits security. Specifically, we used HElayers simulator, which considers the underlying platform capabilities and provides us with the CPU-time of every run, i.e., the needed compute resources for the run. Note that this measurement accumulates the run time of all used CPUs.

Table \ref{tab:acc} summarizes the test accuracy results of four HE-friendly ResNet50 variants: a) a reference HE-friendly ResNet50 network; b) our modified network with shared-source skip connection and Dirac parameterization; c) The reference network without skip connections but with Dirac parameterization; d) The reference network without skip connections. All networks used activation polynomials of degree 8 and were trained over CIFAR-10. For training, we used PyTorch as our library, AdamW as the optimizer (with all default hyperparameters, and learning rate of $1e{-}3$), a batch size of 50, and the standard cross-entropy loss in all our experiments. For simplicity, we did not use dropout or learning rate scheduling. We trained all the networks for 120 epochs and we see that our proposed design provides an interesting tradeoff between CPU-runtime and accuracy. It has almost the same accuracy as the reference implementation but almost the latency of a network without skip connections and an overall CPU-time improvement of around $\times 1.3$. Furthermore, in vanilla networks, the latency and amount of bootstraps operations are proportional to the number of skip-connections. In contrast, in our architecture, these properties are independent of the number of skip-connections, which is crucial for larger networks.

\begin{table}[ht!]
    \centering
    \caption{Comparison of accumulated CPU-time and accuracy for different HE-friendly ResNet50 \glspl{DNN} over CIFAR-10.}
    \label{tab:acc}
    \begin{tabular}{|l|c|c|c|c|}
        \hline
         Network architecture & CPU-time (h) & \# bootstraps & Test accuracy & Test accuracy \\
          &  &  & Non HE-friendly & HE-friendly \\
         \hline
         Reference & 18.06 & 2{,}568 & 91.67 & 91.46 \\
         \hline
         W/o skip connections & 12.9 & 1{,}888 & 88.74 & 88.68\\
         W/ Dirac params & 12.9 & 1{,}888 & 90.87 & 90.74  \\
         \textbf{Our variant} & 13.4 & 1{,}888 & 91.25 & 91.08 \\
         \hline
    \end{tabular}
\end{table}

\begin{figure}[t!]
    \centering
    \includegraphics[width=0.8\textwidth]{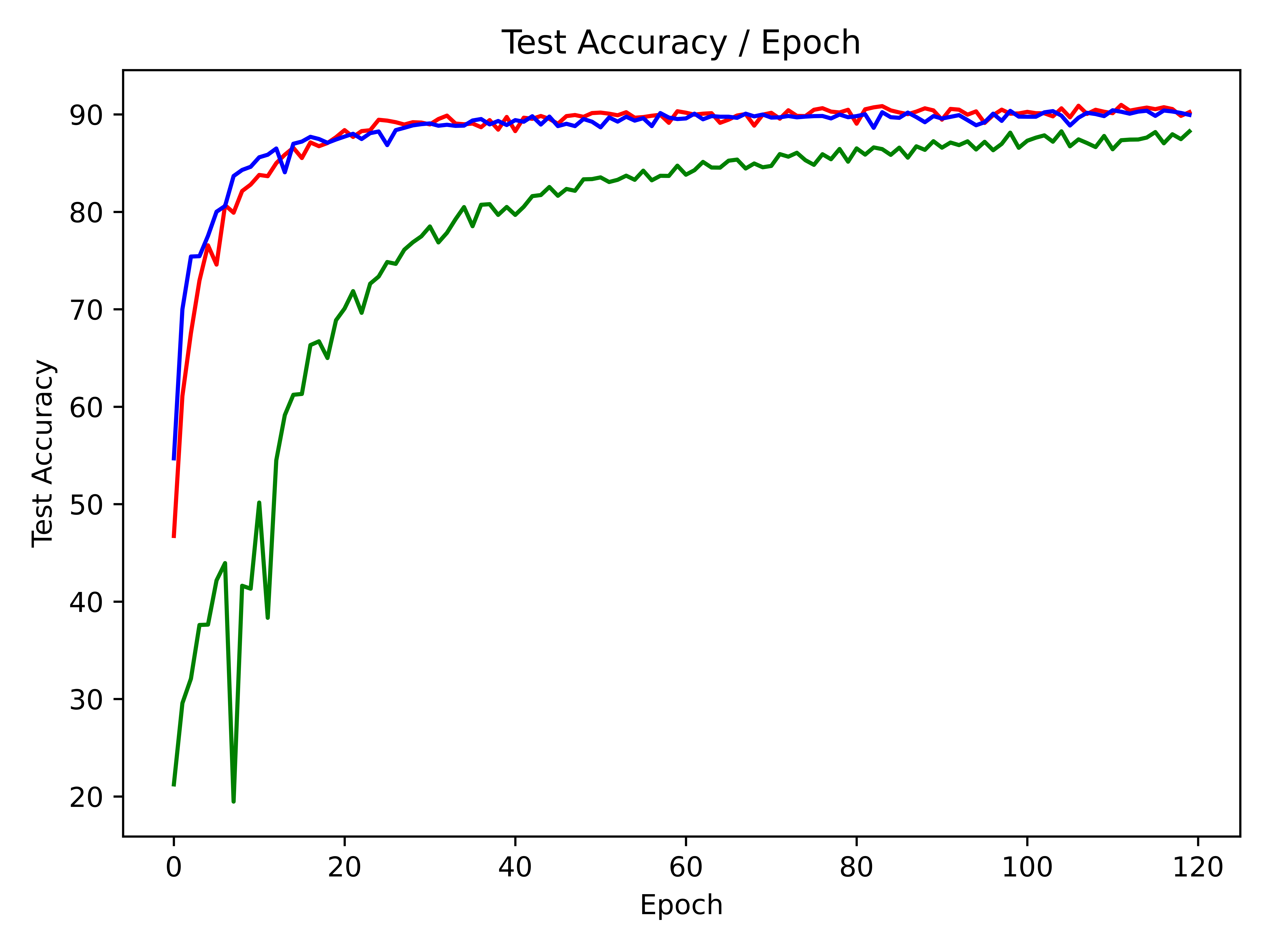}
    \caption{Test accuracy per training epoch of 3 HE-friendly ResNet50 network variants: Reference (blue line), Our modified network (red line), No skip-connection network (green line).}
    \label{fig:accuracy}
\end{figure}

Figure \ref{fig:accuracy} compares the training status of the three HE-friendly ResNet50 variants. For these, it reports the test accuracy (x-axis) per training epoch (y-axis). The reference HE-friendly network is represented by the blue line, our modified network by the red line, and a network where the skip-connections were completely removed by a green line. 

\begin{table}[t!]
    \centering
    \caption{Performance comparison and number of bootstraps used by HElayers to run different HE-friendly ResNet50 \glspl{DNN} over CIFAR-10 with activations approximation of different polynomial degrees.}
    \label{tab:poly}
    \begin{tabular}{|c|l|c|c|}
        \hline
         Activation & Network architecture & CPU-time (h) & \# bootstraps \\
         poly. deg. &  &  & \\
         \hline
         \multirow{3}{*}{2} &
         Reference & 12.33 & $1{,}664$ \\
         & W/o skip connections & 9.13 & 1{,}304 \\
         & \textbf{Our variant} & 10.40 & 1{,}480 \\
         \hline
         \multirow{3}{*}{4} &
         Reference & 11.36 & 1{,}376 \\
         & W/o skip connections & 7.68 & 784 \\
         & \textbf{Our variant} & 8.69 & 784 \\
         \hline
         \multirow{3}{*}{8} &
         Reference & 18.06 & 2{,}568 \\
         & W/o skip connections & 12.92 & 1{,}888 \\
         & \textbf{Our variant} & 13.38 & 1{,}888 \\
         \hline
    \end{tabular}
\end{table}

Table \ref{tab:poly} extends Table \ref{tab:acc} and shows the accumulated CPU-time improvement when using our modified HE-friendly ResNet50 as reported by HElayers simulator \cite{helayers} with activation functions of different polynomial degrees. We observe that compared to the reference network we got an improvement of $ \times 1.18$, $ \times 1.3$, and $ \times 1.34$ in the consumed compute resources when using polynomial activations of degrees $2$, $4$, and $8$, respectively. On the other hand, our network consumed slightly more compute resources compared to a network completely without skip connections, specifically, $\times 0.88$, $\times 0.87$, and $0.97$, for networks with polynomial activations of degrees 2,4, and 8, respectively. Nevertheless, as shown in Figure \ref{fig:accuracy} it achieved the accuracy of the reference implementation. We note that some of the compute resources reduction was due to the reduction in the number of performed bootstraps. However, another reduction comes from the more effective utilization of tile tensors by avoiding several reshaping operations when using shared-source connections.

\section{Conclusions}\label{sec:conc}
Reducing the latency gap between evaluating \glspl{DNN} under HE and in cleartext is of great interest because it allows many organizations to scale their computation and use the cloud when dealing with sensitive data. Our experiments show that it is possible to combine research that originally targeted the AI domain to achieve speedups in a second research domain, namely the security domain. Specifically, we showed that by using techniques such as shared-source skip connections and Dirac parameterization we can reduce the latency of performing inference operations over ResNet50 and CIFAR-10 by $\times 1.3$ while maintaining state-of-the-art accuracy. We believe that this research may open the door to further optimizations that eventually make HE more than practical.

\bibliography{main} 
\bibliographystyle{splncs04}

\newpage
\appendix

\end{document}

%% file: macros.tex
\usepackage{makeidx}  % allows for indexgeneration
\usepackage{graphicx, amssymb}
\usepackage{color}
\usepackage{placeins}
\usepackage{float}
\usepackage{tabularx,colortbl}
\usepackage[svgnames]{xcolor}
\usepackage{mathtools}
\usepackage{cite}
\usepackage{framed}
\usepackage{amsmath,amsfonts} % Math packages
\usepackage{url}
\usepackage{breakurl} 
\usepackage[breaklinks]{hyperref}
\usepackage{comment}
\usepackage{tabularx}
\usepackage{subcaption}
\usepackage{xspace}
\usepackage{tikz-qtree}
\usepackage{multirow}

\hypersetup{
  colorlinks   = true, %Colours links instead of ugly boxes
  urlcolor     = blue, %Colour for external hyperlinks
  linkcolor    = black, %Colour of internal links
  citecolor   = red %Colour of citations
}

\usepackage[nonumberlist,acronyms,nogroupskip]{glossaries}

\newglossary*{not}{Notation}
\usepackage{doi}

\makenoidxglossaries
\setacronymstyle{long-short}

\usepackage{xspace}
\usepackage{commath}
\usepackage[export]{adjustbox}

\usepackage{algorithm}
\usepackage[noend]{algpseudocode}

\usepackage{etoolbox}\AtBeginEnvironment{algorithmic}{\footnotesize}

\usepackage{listings}

\definecolor{codegreen}{rgb}{0,0.6,0}
\definecolor{codegray}{rgb}{0.5,0.5,0.5}
\definecolor{codepurple}{rgb}{0.58,0,0.82}
\definecolor{backcolour}{rgb}{0.95,0.95,0.92}

\definecolor{gray}{gray}{0.5}
\colorlet{commentcolour}{green!50!black}

\colorlet{stringcolour}{red!60!black}
\colorlet{keywordcolour}{magenta!90!black}
\colorlet{exceptioncolour}{yellow!50!red}
\colorlet{commandcolour}{blue!60!black}
\colorlet{numpycolour}{blue!60!green}
\colorlet{literatecolour}{magenta!90!black}
\colorlet{promptcolour}{green!50!black}
\colorlet{specmethodcolour}{violet}

\definecolor{lgrey}{rgb}{0.9,0.9,0.9}
\definecolor{asparagus}{rgb}{0.53, 0.66, 0.42}

\lstdefinelanguage
   [x64]{Assembler}     % add a "x64" dialect of Assembler
   [x86masm]{Assembler} % based on the "x86masm" dialect
   {morekeywords={VMOVDQA64, VPERMPD, VXORPD, VPCLMULQDQ, %
                  VMOVDQU64, VALIGNQ, VAESENC, VAESENCLAST, 
                  VAESDEC, VAESDECLAST, VBROADCASTI64X2, VPXORQ, 
                  VPXOR, VMOVDQU, VPADD, vpsrldq, vpslldq, 
                  vpshufd,vextracti64x4, vextracti128, vpermb,
                  vpblendvb, aesenc, pshufb, zmm0, xmm, rax, rdi, retq, nopw, rcx,
                  .byte,
                  .long, .set},
    morecomment=[l][\color{asparagus}]{\#}} % etc.

\lstdefinestyle{mystyle}{
    backgroundcolor=\color{lgrey},  
      basicstyle=\scriptsize \ttfamily \color{black} \bfseries,
      breakatwhitespace=false,
      breaklines=true,
      captionpos=b,
      commentstyle=\color{asparagus},
      deletekeywords={...},
      escapeinside={\%*}{*)},
      frame=single,
      keywordstyle=\color{purple},
      identifierstyle=\color{black},
      stringstyle=\color{blue},
      language=[x64]Assembler,
      numbers=left,
      numbersep=5pt,
      numberstyle=\scriptsize\color{black},
      rulecolor=\color{black},
      showspaces=false,
      showstringspaces=false,
      showtabs=false,
      stepnumber=1,
      tabsize=5
}
 
\lstdefinestyle{mystyleC}{
    backgroundcolor=\color{lgrey},  
      basicstyle=\scriptsize \ttfamily \color{black} \bfseries,
      breakatwhitespace=false,
      breaklines=true,
      captionpos=b,
      commentstyle=\color{asparagus},
      deletekeywords={...},
      escapeinside={\%*}{*)},
      frame=single,
      keywordstyle=\color{purple},
      identifierstyle=\color{black},
      stringstyle=\color{blue},
      language=C,
      numbers=left,
      numbersep=5pt,
      numberstyle=\scriptsize\color{black},
      rulecolor=\color{black},
      showspaces=false,
      showstringspaces=false,
      showtabs=false,
      stepnumber=1,
      tabsize=5
}

\lstdefinestyle{mystyleP}{
    backgroundcolor=\color{lgrey},  
      basicstyle=\footnotesize \ttfamily \color{black} \bfseries,
      breakatwhitespace=false,
      breaklines=true,
      captionpos=b,
      commentstyle=\color{asparagus},
      deletekeywords={...},
      escapeinside={\%*}{*)},
      frame=single,
      keywordstyle=\color{purple},
      morekeywords={},
      identifierstyle=\color{black},
      stringstyle=\color{blue},
      language=Python,
      numbers=none,
      numbersep=5pt,
      numberstyle=\tiny\color{black},
      rulecolor=\color{black},
      showspaces=false,
      showstringspaces=false,
      showtabs=false,
      stepnumber=1,
      tabsize=5,
      title=\lstname
}

\lstnewenvironment{asmcode}[1][]%
  {\minipage{0.96\linewidth}\medskip 
   \lstset{style=mystyle,frame=single,#1}}
  {\endminipage}

\lstnewenvironment{ccode}[1][]%
  {\minipage{0.96\linewidth}\medskip 
   \lstset{style=mystyleC,frame=single,#1}}
  {\endminipage}

\lstnewenvironment{pythoncode}[1][]%
  {\minipage{0.96\linewidth}\medskip 
   \lstset{style=mystyleP,frame=single,#1}}
  {\endminipage}

\newtheorem{observation}{Observation}

\newcommand{\circler}{\textsuperscript{\textregistered}}

\newcommand{\R}{\mathcal{R}}

\newcommand{\Math}[1]{\ensuremath{#1}\xspace}

\newcommand{\Enc}{\Math{\operatorname{Enc}}}
\newcommand{\Dec}{\Math{\operatorname{Dec}}}
\newcommand{\ChainIdx}{\Math{\operatorname{CIdx}}}
\newcommand{\ReScale}{\Math{\operatorname{ReScale}}}
\newcommand{\Bootstrap}{\Math{\operatorname{Bootstrap}}}

\renewcommand{\paragraph}[1]{{\medskip\noindent {\bf{#1}}~}}

\usepackage{tikz}
\usetikzlibrary{svg.path}
\definecolor{orcidlogocol}{HTML}{A6CE39}
\tikzset{
  orcidlogo/.pic={
    \fill[orcidlogocol] svg{M256,128c0,70.7-57.3,128-128,128C57.3,256,0,198.7,0,128C0,57.3,57.3,0,128,0C198.7,0,256,57.3,256,128z};
    \fill[white] svg{M86.3,186.2H70.9V79.1h15.4v48.4V186.2z}
                 svg{M108.9,79.1h41.6c39.6,0,57,28.3,57,53.6c0,27.5-21.5,53.6-56.8,53.6h-41.8V79.1z M124.3,172.4h24.5c34.9,0,42.9-26.5,42.9-39.7c0-21.5-13.7-39.7-43.7-39.7h-23.7V172.4z}
                 svg{M88.7,56.8c0,5.5-4.5,10.1-10.1,10.1c-5.6,0-10.1-4.6-10.1-10.1c0-5.6,4.5-10.1,10.1-10.1C84.2,46.7,88.7,51.3,88.7,56.8z};
  }
}
\newcommand{\OrigHeightRecip}{0.00390625}
\newlength{\curXheight}
\DeclareRobustCommand\orcid[1]{%
\texorpdfstring{%
\setlength{\curXheight}{\fontcharht\font`X}%
\href{https://orcid.org/#1}{\XeTeXLinkBox{\mbox{%
\begin{tikzpicture}[yscale=-\OrigHeightRecip*\curXheight,
xscale=\OrigHeightRecip*\curXheight,transform shape]
\pic{orcidlogo};
\end{tikzpicture}%
}}}}{}}
\newcommand{\todo}[1][]{\textcolor{red}}

%% file: _header.tex
% \title{Efficient skip connections Realization For Secure Inference on Encrypted Data with highway and Dirac Connections}
\title{Efficient Skip Connections Realization for Secure Inference on Encrypted Data
%with {\color{red}Shared-source skip connection and Dirac Parameterization }
}
%\subtitle{-- DRAFT -- CONFIDENTIAL --}

%\authorrunning{}
\titlerunning{Skip Connections Realization for Secure Inference}
\authorrunning{Nir Drucker, Itamar Zimerman}
%\titlerunning{ResNet under Encrpytion}

\author{Nir Drucker\orcid{0000-0002-7273-4797} \and
Itamar Zimerman\orcid{0000-0001-8321-0609}}

\institute{IBM Research - Israel} 

%\author{Anonymized for the review}
%\institute{Confidential}

\maketitle

%% file: _abstract.tex
\begin{abstract}
Homomorphic Encryption (HE) is a cryptographic tool that allows performing computation under encryption, which is used by many privacy-preserving machine learning solutions, for example, to perform secure classification. 
Modern deep learning applications yield good performance for example in image processing tasks benchmarks by including many skip connections. The latter appears to be very costly when attempting to execute model inference under HE. 
In this paper, we show that by replacing (mid-term) skip connections with (short-term) Dirac parameterization and (long-term) shared-source skip connection we were able to reduce the skip connections burden for HE-based solutions, achieving $\times 1.3$ computing power improvement for the same accuracy.
\end{abstract}

\keywords{shared-source skip connections, Dirac networks, Dirac parameterization, homomorphic encryption, privacy preserving machine learning, PPML, encrypted neural networks, deep neural networks}

%% file: main.bbl
\begin{thebibliography}{10}
\providecommand{\url}[1]{\texttt{#1}}
\providecommand{\urlprefix}{URL }
\providecommand{\doi}[1]{https://doi.org/#1}

\bibitem{helayers}
Aharoni, E., Adir, A., Baruch, M., Drucker, N., Ezov, G., Farkash, A.,
  Greenberg, L., Masalha, R., Moshkowich, G., Murik, D., et~al.: {HElayers: A
  tile tensors framework for large neural networks on encrypted data}. PoPETs
  (2023). \doi{10.56553/popets-2023-0020}

\bibitem{standard}
Albrecht, M., Chase, M., Chen, H., Ding, J., Goldwasser, S., Gorbunov, S.,
  Halevi, S., Hoffstein, J., Laine, K., Lauter, K., Lokam, S., Micciancio, D.,
  Moody, D., Morrison, T., Sahai, A., Vaikuntanathan, V.: Homomorphic
  encryption security standard. Tech. rep., HomomorphicEncryption.org, Toronto,
  Canada (November 2018), \url{https://HomomorphicEncryption.org}

\bibitem{SP3}
Balduzzi, D., Frean, M., Leary, L., Lewis, J.P., Ma, K.W.D., McWilliams, B.:
  The shattered gradients problem: If resnets are the answer, then what is the
  question? In: Precup, D., Teh, Y.W. (eds.) Proceedings of the 34th
  International Conference on Machine Learning. Proceedings of Machine Learning
  Research, vol.~70, pp. 342--350. PMLR (06--11 Aug 2017),
  \url{https://proceedings.mlr.press/v70/balduzzi17b.html}

\bibitem{e2e}
Baruch, M., Drucker, N., Ezov, G., Kushnir, E., Lerner, J., Soceanu, O.,
  Zimerman, I.: {Sensitive Tuning of Large Scale CNNs for E2E Secure Prediction
  using Homomorphic Encryption} (2023), \url{https://arxiv.org/abs/2304.14836}

\bibitem{HIPAA}
{Centers for Medicare \& Medicaid Services}: {The Health Insurance Portability
  and Accountability Act of 1996 (HIPAA)} (1996),
  \url{https://www.hhs.gov/hipaa/}

\bibitem{ckks-rns}
Cheon, J.H., Han, K., Kim, A., Kim, M., Song, Y.: {A Full RNS Variant of
  Approximate Homomorphic Encryption}. In: Cid, C., {Jacobson Jr.}, M.J. (eds.)
  Selected Areas in Cryptography -- SAC 2018. pp. 347--368. Springer
  International Publishing, Cham (2019). \doi{10.1007/978-3-030-10970-7\_16}

\bibitem{ckks2017}
Cheon, J.H., Kim, A., Kim, M., Song, Y.: Homomorphic encryption for arithmetic
  of approximate numbers. In: International Conference on the Theory and
  Application of Cryptology and Information Security. pp. 409--437. Springer
  (2017). \doi{10.1007/978-3-319-70694-8\_15}

\bibitem{heaanCode}
CryptoLab: {HEaaN: Homomorphic Encryption for Arithmetic of Approximate
  Numbers} (2022), \url{https://www.cryptolab.co.kr/eng/product/heaan.php}

\bibitem{dosovitskiy2020image}
Dosovitskiy, A., Beyer, L., Kolesnikov, A., Weissenborn, D., Zhai, X.,
  Unterthiner, T., Dehghani, M., Minderer, M., Heigold, G., Gelly, S., et~al.:
  An image is worth 16x16 words: Transformers for image recognition at scale.
  arXiv preprint arXiv:2010.11929  (2020),
  \url{https://arxiv.org/abs/2010.11929}

\bibitem{GDPR}
{EU General Data Protection Regulation}: Regulation ({EU}) 2016/679 of the
  {E}uropean {P}arliament and of the {C}ouncil of 27 {A}pril 2016 on the
  protection of natural persons with regard to the processing of personal data
  and on the free movement of such data, and repealing {D}irective {95/46/EC}
  ({G}eneral {D}ata {P}rotection {R}egulation). Official Journal of the
  European Union  \textbf{119} (2016),
  \url{http://data.europa.eu/eli/reg/2016/679/oj}

\bibitem{gartner}
Gartner: Gartner identifies top security and risk management trends for 2021.
  Tech. rep. (March 2021),
  \url{https://www.gartner.com/en/newsroom/press-releases/2021-03-23-gartner-identifies-top-security-and-risk-management-t}

\bibitem{CryptoNets2016}
Gilad~Bachrach, R., Dowlin, N., Laine, K., Lauter, K., Naehrig, M., Wernsing,
  J.: Cryptonets: Applying neural networks to encrypted data with high
  throughput and accuracy. In: International Conference on Machine Learning.
  pp. 201--210 (2016),
  \url{http://proceedings.mlr.press/v48/gilad-bachrach16.pdf}

\bibitem{Halevi2017}
Halevi, S.: {Homomorphic Encryption}. In: Lindell, Y. (ed.) Tutorials on the
  Foundations of Cryptography: Dedicated to Oded Goldreich, pp. 219--276.
  Springer International Publishing, Cham (2017).
  \doi{10.1007/978-3-319-57048-8\_5}

\bibitem{sm}
He, K., Zhang, X., Ren, S., Sun, J.: Deep residual learning for image
  recognition. In: Proceedings of the IEEE Conference on Computer Vision and
  Pattern Recognition (CVPR) (June 2016),
  \url{https://openaccess.thecvf.com/content_cvpr_2016/html/He_Deep_Residual_Learning_CVPR_2016_paper.html}

\bibitem{pmlr-v162-lee22e}
Lee, E., Lee, J.W., Lee, J., Kim, Y.S., Kim, Y., No, J.S., Choi, W.:
  Low-complexity deep convolutional neural networks on fully homomorphic
  encryption using multiplexed parallel convolutions. In: Chaudhuri, K.,
  Jegelka, S., Song, L., Szepesvari, C., Niu, G., Sabato, S. (eds.) Proceedings
  of the 39th International Conference on Machine Learning. vol.~162, pp.
  12403--12422. PMLR (17--23 Jul 2022),
  \url{https://proceedings.mlr.press/v162/lee22e.html}

\bibitem{liu2022convnet}
Liu, Z., Mao, H., Wu, C.Y., Feichtenhofer, C., Darrell, T., Xie, S.: A convnet
  for the 2020s. In: Proceedings of the IEEE/CVF Conference on Computer Vision
  and Pattern Recognition. pp. 11976--11986 (2022)

\bibitem{oord2016wavenet}
van~den Oord, A., Dieleman, S., Zen, H., Simonyan, K., Vinyals, O., Graves, A.,
  Kalchbrenner, N., Senior, A., Kavukcuoglu, K.: Wavenet: A generative model
  for raw audio. In: 9th ISCA Speech Synthesis Workshop. pp. 125--125 (2016),
  \url{https://www.isca-speech.org/archive_v0/SSW_2016/abstracts/ssw9_DS-4_van_den_Oord.html}

\bibitem{NS2}
Oyedotun, O.K., Shabayek, A.E.R., Aouada, D., Ottersten, B.: Going deeper with
  neural networks without skip connections. In: 2020 IEEE International
  Conference on Image Processing (ICIP). pp. 1756--1760 (2020).
  \doi{10.1109/ICIP40778.2020.9191356}

\bibitem{sp2}
Pascanu, R., Mikolov, T., Bengio, Y.: On the difficulty of training recurrent
  neural networks. In: Dasgupta, S., McAllester, D. (eds.) Proceedings of the
  30th International Conference on Machine Learning. Proceedings of Machine
  Learning Research, vol.~28, pp. 1310--1318. PMLR, Atlanta, Georgia, USA
  (17--19 Jun 2013), \url{https://proceedings.mlr.press/v28/pascanu13.html}

\bibitem{GPT}
Radford, A., Narasimhan, K., Salimans, T., Sutskever, I., et~al.: Improving
  language understanding by generative pre-training  (2018),
  \url{https://www.cs.ubc.ca/~amuham01/LING530/papers/radford2018improving.pdf}

\bibitem{new-highway}
Tai, Y., Yang, J., Liu, X.: Image super-resolution via deep recursive residual
  network. In: 2017 IEEE Conference on Computer Vision and Pattern Recognition
  (CVPR). pp. 2790--2798 (2017). \doi{10.1109/CVPR.2017.298}

\bibitem{hebench}
{The HEBench Organization}: {HEBench} (2022), \url{https://hebench.github.io/}

\bibitem{vaswani2017attention}
Vaswani, A., Shazeer, N., Parmar, N., Uszkoreit, J., Jones, L., Gomez, A.N.,
  Kaiser, {\L}., Polosukhin, I.: Attention is all you need. Advances in neural
  information processing systems  \textbf{30} (2017),
  \url{https://proceedings.neurips.cc/paper/2017/file/3f5ee243547dee91fbd053c1c4a845aa-Paper.pdf}

\bibitem{diracnet}
Zagoruyko, S., Komodakis, N.: Diracnets: Training very deep neural networks
  without skip-connections (2017), \url{https://arxiv.org/abs/1706.00388}

\bibitem{NS1}
Zagoruyko, S., Komodakis, N.: Diracnets: Training very deep neural networks
  without skip-connections. CoRR  \textbf{abs/1706.00388} (2017),
  \url{http://arxiv.org/abs/1706.00388}

\end{thebibliography}
